\documentclass[aps,preprint]{revtex4}%
\usepackage{amsfonts}
\usepackage{amsmath}
\usepackage{amssymb}
\usepackage{graphicx}%
\begin{document}
\preprint{ }
\title[Altered Maxwell equations]{Altered Maxwell equations in the length gauge}
\author{H. R. Reiss}
\affiliation{Max Born Institute, 12489 Berlin, Germany}
\affiliation{American University, Washington, DC 20016-8058, USA}


\begin{abstract}
The length gauge uses a scalar potential to describe a laser field, thus
treating it as a longitudinal field rather than as a transverse field. This
distinction is revealed in the fact that the Maxwell equations that relate to
the length gauge are not the same as those for transverse fields. In
particular, a source term is necessary in the length-gauge Maxwell equations,
whereas the Coulomb-gauge description of plane waves possesses the basic
property of transverse fields that they propagate with no source terms at all.
This difference is shown to be importantly consequential in some previously
unremarked circumstances; and it explains why the G\"{o}ppert-Mayer gauge
transformation does not provide the security that might be expected of full
gauge equivalence.

\end{abstract}
\date[7 April 2013]{}
\maketitle

\section{Introduction}

It is shown here that the \textit{length gauge} (LG) is based on a different
set of Maxwell equations than those that relate to plane-wave fields. These
differences can generate errors, for example, in applications of the
\textquotedblleft simpleman\textquotedblright\ method\cite{corkum}. It is also
shown that the squared vector potential $\mathbf{A}^{2}\left(  t\right)  $ for
a plane wave field in the dipole approximation, cannot be transformed away
despite the usual rule about quantities in the Hamiltonian that depend only on
the time. These problems arise from a excessively strong form of the dipole
approximation, necessary in the LG, that requires the complete absence of a
magnetic field and the restriction of electric fields to time dependence
alone. Working within the \textit{Coulomb gauge }(CG), a much milder form of
the dipole approximation can be used that makes possible some simplifications
(but only when rigorously applicable) while still retaining the proper Maxwell
equations. Employment of the LG for laser problems is equivalent to the
treatment of plane-wave fields as if they were quasistatic electric (QSE)
fields. An important distinction between plane-wave fields and QSE fields is
that plane-wave fields have the unique property of propagation in the absence
of sources. In contrast to this feature, QSE fields require sources, mandating
the existence of a \textquotedblleft virtual\textquotedblright\ source in the
LG, that is not actually present in the laboratory. It is shown that this
virtual source can introduce unphysical anomalies when using the
G\"{o}ppert-Mayer gauge transformation\cite{gm} to introduce the LG.

For single-active-electron atomic ionization problems, the LG interaction
Hamiltonian has the form%
\begin{equation}
H_{I}^{LG}=\mathbf{r\cdot E}\left(  t\right)  , \label{a}%
\end{equation}
and the CG interaction Hamiltonian is%
\begin{equation}
H_{I}^{CG}=\frac{1}{c}\widehat{\mathbf{p}}\mathbf{\cdot A}\left(
t,\mathbf{r}\right)  +\frac{1}{2c^{2}}\mathbf{A}^{2}\left(  t,\mathbf{r}%
\right)  , \label{b}%
\end{equation}
where $\widehat{\mathbf{p}}$ is the momentum operator, $\mathbf{E}\left(
t\right)  $ is the dipole-approximation electric field, and $\mathbf{A}\left(
t,\mathbf{r}\right)  $ is a fully-stated vector potential. Atomic units are
used, and electromagnetic quantities are in Gaussian units. \textit{Dipole
approximation} is here taken to mean that the phase $\omega t-\mathbf{k\cdot
r}$ of a plane wave ($\mathbf{k}$ is the propagation vector) is approximated
by $\omega t$ alone. This will be referred to as the \textit{strong form }of
the dipole approximation since it has the consequence that%
\begin{equation}
\mathbf{E}=\mathbf{E}\left(  t\right)  ,\qquad\mathbf{B}=0. \label{1a}%
\end{equation}
An important feature of the strong-form dipole approximation is that the
fundamental Lorentz invariant $\mathbf{E}^{2}-\mathbf{B}^{2}$ has a positive
value
\begin{equation}
\left(  \mathbf{E}^{2}-\mathbf{B}^{2}\right)  ^{LG}>0, \label{2a}%
\end{equation}
in the LG rather than the zero value%
\begin{equation}
\left(  \mathbf{E}^{2}-\mathbf{B}^{2}\right)  ^{PW}=0 \label{3a}%
\end{equation}
that characterizes plane-wave (PW) fields. (The descriptions \textquotedblleft
plane-wave field\textquotedblright\ and \textquotedblleft transverse
field\textquotedblright\ will be used interchangeably.) This basic departure
from plane-wave behavior becomes increasingly important as the field strength
increases, thus explaining why this distinction seems not to be a source of
difficulty in the context of conventional perturbative Atomic, Molecular, and
Optical (AMO) physics. It can, however, be critically important in very
intense laser fields\cite{hr101,hr102,hr75}.

The G\"{o}ppert-Mayer\cite{gm} gauge transformation ostensibly connects the LG
and the CG and makes it possible to achieve (in many problems) valid results
in both gauges. The LG is useful because it requires only a scalar potential,
which is generally much easier to employ than is a vector potential. However,
the limitation to a scalar potential has several consequences that are often
overlooked. The $\mathbf{r\cdot E}\left(  t\right)  $ interaction Hamiltonian
(\ref{a}) is exactly the same as pertains to the presence of a quasistatic
electric (QSE) field. The limitation to a scalar potential means that a
propagating solution does not exist at all. A QSE field is a longitudinal
field; an entity that is fundamentally different from the transverse field
that it is intended to mimic. A QSE field oscillates with time, but it does
not propagate through space with time in the manner of a plane-wave field.
\textit{There is no solution of the Maxwell equations that describes a
propagating field when only a scalar potential exists.} The G\"{o}ppert-Mayer
(GM) gauge transformation essentially amounts to the use of a longitudinal
field as a proxy for a transverse field.

In her 1931 paper\cite{gm}, G\"{o}ppert-Mayer remarks that the wavelength of
the field should be much larger than the size of the atomic system with which
it interacts. This is not a limitation that can be deduced from the LG itself;
rather, it comes from the fact that G\"{o}ppert-Mayer was aware that she was
using an approximation to a propagating field, and that this approximation is
not applicable if the wavelength of the propagating field is short enough to
probe\ the details of the atomic structure. However, she makes no mention of a
limitation at low frequencies. That omission remains largely in place at
present, despite the fact that, in the\ modern laser environment, it is the
low frequency limitation that is of far greater importance\cite{hr101,hr102}
than the high frequency limit.

One of the most important conclusions reached below, however, is that
gauge-related differences exist even within the parameter space in which the
dipole approximation is nominally valid.

Section II discusses the Lorentz invariants that can be used to identify
classes of electromagnetic phenomena. This is of critical importance because
the LG and the CG are associated with different electromagnetic classes. These
different classes are governed by different sets of Maxwell equations. In
particular, the solution of the classical Maxwell equations employed in the
theory of rescattering phenomena\cite{corkum} does not relate to underlying
properties of plane waves, but rather to the often quite different properties
of quasistatic electric fields.

A vital element in the discussion of relevant Maxwell equations arises from
the realization that a description of laser-induced phenomena in the LG
requires the introduction of a virtual source current that is not actually
present in the laboratory. In some circumstances, this virtual current can
introduce unphysical energy or angular momentum into a problem.

A basic imperative that emerges in this section on electrodynamics is the
necessity to retain the squared vector potential $\mathbf{A}^{2}$ even when
the dipole approximation appears to make it a function only of time. The
familiar rule that an additive quantity in the Hamiltonian or Lagrangian
function plays no role in the equations of motion is not applicable when that
term is large and arises when using an approximate treatment of plane wave fields.

The preservation of basic CG properties even when dipole-approximation forms
are justified is verified by formulating problems such as the Strong-Field
Approximation (SFA) in a fully relativistic context\cite{hr90,hrrev}, and then
demonstrating that the dipole-approximation result follows as a limit of the
fully relativistic case\cite{hrrel,craw1,craw2}. The results support the need
to retain $\mathbf{A}^{2}$ when it is of significant magnitude.

Section III examines those problems where these basic differences in the LG
and the CG can cause major errors to occur. For example, the simpleman
method\cite{corkum} can give erroneous predictions with circularly polarized
fields even within the domain of field parameters where the dipole
approximation is justified in the CG. Problems also arise in the significantly
different ranges of applicability of the Strong Field Approximation (SFA) in
the LG and in the CG. For example, the SFA in the dipole-approximated CG can
give excellent results for plane-wave fields of quite high
frequency\cite{bondar,hrhfa,hropex}, whereas the LG leads to tunneling models
that have no applicability at all for high frequencies\cite{hr101}.

Section IV discusses the limitations in the LG description of plane-wave
phenomena when the putative Volkov solution (an explicitly propagating-wave
solution) is employed in a gauge that does not possess a traveling-wave property.

Finally, Section V summarizes the essential findings of the paper.

\section{Basic electrodynamic background}

The characteristics of an electromagnetic field inferred from the Lorentz
invariants associated with different field configurations are reviewed, since
they serve as a prelude to examining the Maxwell equations directly. Gaussian
units are used to achieve clarity through simplicity of notation.

\subsection{Lorentz invariants}

The basic Lorentz invariants that characterize an electromagnetic field are
inferred from the field tensor, which is%
\begin{equation}
F^{\mu\nu}=\partial^{\mu}A^{\nu}-\partial^{\nu}A^{\mu}, \label{ca}%
\end{equation}
where $A^{\mu}$ is the 4-vector potential that describes the electromagnetic
field. The relativistic conventions employed are those of
Jackson\cite{jackson}. A dual tensor $\mathcal{F}^{\mu\nu}$ can also be
defined by acting on the field tensor with the completely antisymmetric
4-space tensor $\epsilon^{\mu\nu\rho\lambda}$. The two Lorentz invariants that
can be formed from scalar products of these tensors are\cite{jackson}%
\begin{align}
F^{\mu\nu}F_{\mu\nu}  &  =-2\left(  \mathbf{E}^{2}-\mathbf{B}^{2}\right)
,\label{d}\\
\mathcal{F}^{\mu\nu}F_{\mu\nu}  &  =-4\mathbf{E\cdot B}, \label{e}%
\end{align}
often referred to simply as the $\mathbf{E}^{2}-\mathbf{B}^{2}$ and
$\mathbf{E\cdot B}$ scalar Lorentz invariants. They make possible the simple
chart%
\begin{equation}%
\begin{tabular}
[c]{|c||c|c|}\hline
& $\mathbf{E}^{2}-\mathbf{B}^{2}$ & $\mathbf{E\cdot B}$\\\hline\hline
constant electric field & $\mathbf{E}^{2}$ & $0$\\\hline
plane wave field & $0$ & $0$\\\hline
constant magnetic field & $-\mathbf{B}^{2}$ & $0$\\\hline
\end{tabular}
\ \ \ \ \ \ \ \ \ \label{f}%
\end{equation}

The $\mathbf{E\cdot B}$ invariant is zero for all three cases, albeit for
different reasons. It is the $\mathbf{E}^{2}-\mathbf{B}^{2}$ invariant that
makes the important distinctions: it is positive-definite for the constant
electric field (and the QSE field), negative definite for the constant
magnetic field, and zero for the plane wave field. The stronger the field, the
more important the distinctions become. The plane wave field is a
\textit{transverse field} that has the properties%
\begin{equation}
\left\vert \mathbf{E}\right\vert =\left\vert \mathbf{B}\right\vert
,\quad\mathbf{E\bot B} \label{g}%
\end{equation}
that underlie the entries in (\ref{f}). It is called transverse because the
electric and magnetic fields are both perpendicular to the direction of
propagation of the field as well as to each other. The constant electric field
is the limiting case of a QSE field or \textit{longitudinal field}, where the
sole preferred direction in the physical problem is the direction of the
electric field. From the chart (\ref{f}) it can be said that the constant
electric field is as different from a plane wave as is the constant magnetic field.

\textit{It is emphasized that the basic properties of a plane wave field as
expressed in }(\ref{f}) \textit{and} (\ref{g})\textit{ cannot be replicated by
a scalar potential.}

\subsection{Maxwell equations}

The Maxwell equations for the vacuum are%
\begin{align}
\mathbf{\nabla\cdot E}  &  =4\pi\rho,\label{h}\\
\mathbf{\nabla\cdot B}  &  =0,\label{i}\\
\mathbf{\nabla\times B}-\frac{1}{c}\frac{\partial}{\partial t}\mathbf{E}  &
=\frac{4\pi}{c}\mathbf{J},\label{j}\\
\mathbf{\nabla\times E}+\frac{1}{c}\frac{\partial}{\partial t}\mathbf{B}  &
=0, \label{k}%
\end{align}
where $\rho$ and $\mathbf{J}$ are source terms, being respectively a charge
density and a current density.

\subsection{Maxwell equations for plane waves}

Plane waves have the distinction that, once formed, they can propagate without
limit in the complete absence of sources. The Maxwell equations appropriate to
plane waves are found from Eqs. (\ref{h}) - (\ref{k}) by giving zero values to
the sources. This gives the Maxwell equations for plane waves as%
\begin{align}
\mathbf{\nabla\cdot E}  &  =0,\label{l}\\
\mathbf{\nabla\cdot B}  &  =0,\label{m}\\
\mathbf{\nabla\times B}-\frac{1}{c}\frac{\partial}{\partial t}\mathbf{E}  &
=0,\label{n}\\
\mathbf{\nabla\times E}+\frac{1}{c}\frac{\partial}{\partial t}\mathbf{B}  &
=0. \label{o}%
\end{align}

\subsection{Dipole approximation}

When plane waves interact with a bound system whose size is much smaller than
a wavelength $\lambda$ of the field, this sets an upper limit on the frequency
for which the dipole approximation is applicable, given by%
\[
\lambda\gg a_{0},\text{ \ or \ }\omega\ll\frac{2\pi c}{a_{0}},
\]
where $a_{0}$, the Bohr radius, characterizes the size of an atom. In atomic
units (a.u.), used henceforth, this is just%
\begin{equation}
\omega\ll2\pi c. \label{o1}%
\end{equation}
A lower limit on the frequencies for which the dipole approximation can be
applied comes from the onset of specific effects of the magnetic field, given
by\cite{hrrev,hr101}%
\begin{equation}
\omega\gg\frac{1}{2}\left(  \frac{I}{c}\right)  ^{1/3} \label{o2}%
\end{equation}
for linear polarization, where $I$ is the field intensity. This lower
frequency limit on the dipole approximation does not occur naturally in the
LG, where it is often overlooked.

G\"{o}ppert-Mayer, in her famous 1931 paper\cite{gm}, specified only the upper
limit on the frequency that is given by Eq. (\ref{o1}). This overlooking of
the need to fully consider the requirements for complete neglect of the
magnetic field is not surprising for a 1931 paper. However, that 80-year-old
oversight continues to prevail in almost all current work.

The dipole approximation is often taken to be characterized by the application
of the strong form as given in Eq. (\ref{1a}). These conditions are not
possible in terms of Eqs. (\ref{l}) - (\ref{o}). This means that the dipole
approximation for a plane-wave field cannot be stated simply as Eq. (\ref{1a}).

It must be kept in mind that plane waves are unique in that they are the only
propagating-wave solution of the Maxwell equations. Transverse fields cannot
propagate unless the magnetic field is present and equal in magnitude to the
electric field. The conditions (\ref{1a}) are an over-simplification when
applied to plane waves, whereas the properties (\ref{g}) are \textit{always}
valid and necessary for plane waves. When a specific quantum matrix element is
to be evaluated, it may be possible to employ the approximations (\ref{1a})
without affecting the value of the matrix element. It is also true that, even
when $\left\vert \mathbf{B}\right\vert =\left\vert \mathbf{E}\right\vert $,
the classical force exerted by the magnetic field on a charged particle can be
negligible compared to the electric force because of the Lorentz force%
\[
\mathbf{F}=q\left(  \mathbf{E}+\frac{\mathbf{v}}{c}\times\mathbf{B}\right)
\]
exerted on a particle of charge $q$. Nevertheless, plane waves are always
associated with zero values for the $\mathbf{E}^{2}-\mathbf{B}^{2}$ and
$\mathbf{E\cdot B}$ invariants listed in the chart (\ref{f}). The upper limit
on the frequency for which the conditions (\ref{1a}) are harmless
approximations for plane wave phenomena is given by Eq. (\ref{o1}), and the
lower limit comes from Eq. (\ref{o2}). Within those bounds, the actual
presence of a magnetic field with $\left\vert \mathbf{B}\right\vert
=\left\vert \mathbf{E}\right\vert $ does not have major consequences for
linear polarization problems beyond that of maintaining the propagation
property of a plane wave. The circular polarization case (and thus elliptical
polarization approaching circular) involves more restrictive requirements to
be examined below.

\subsection{Maxwell equations for the length gauge}

The length gauge explicitly obeys the conditions in Eq. (\ref{1a}). To state
the appropriate Maxwell equations, it is necessary to return to the complete
expression of the Maxwell equations in Eqs. (\ref{h}) - (\ref{k}). The
conditions (\ref{1a}) reduce the four Maxwell equations (\ref{h}) - (\ref{k})
to the single equation%
\begin{equation}
\frac{\partial}{\partial t}\mathbf{E}\left(  t\right)  =-4\pi J_{v}. \label{r}%
\end{equation}
The subscript $v$ has been appended to the current term to indicate that it is
a virtual current that does not really exist in the laboratory; it is
necessary in order to support a non-trivial result for the behavior of a
charged particle described in the length gauge.

\begin{figure}
[ptb]
\begin{center}
\includegraphics[
height=4.561in,
width=5.8781in
]%
{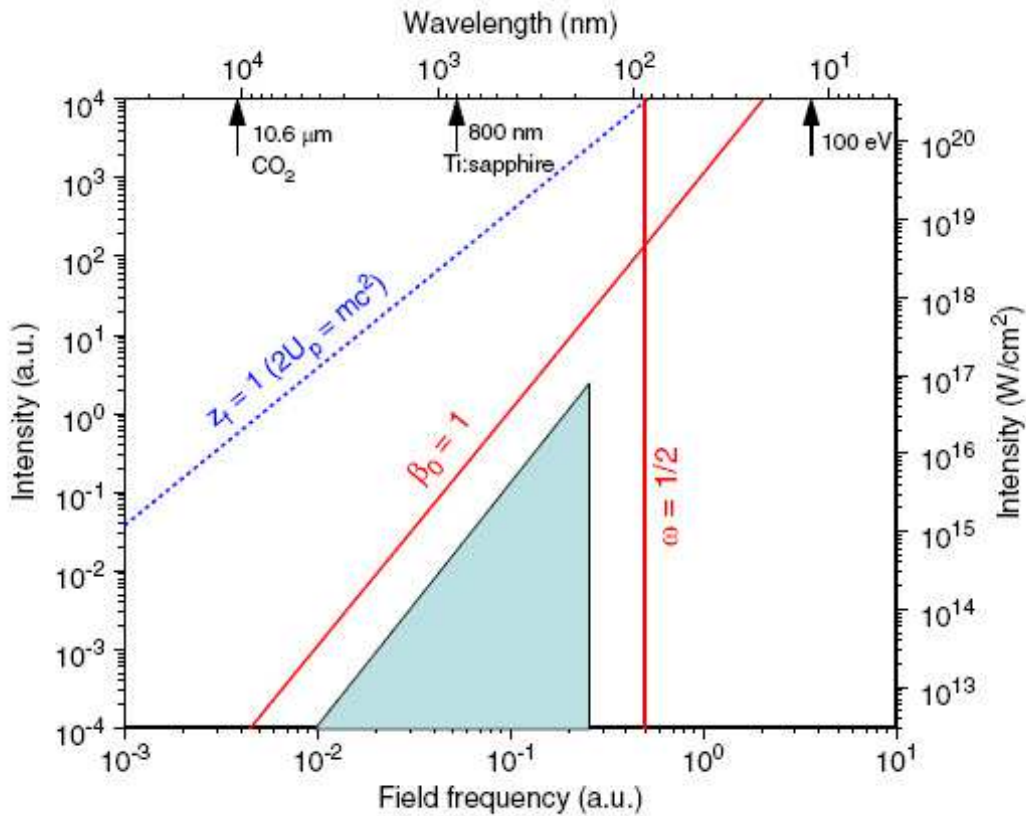}%
\caption{This figure, taken from Ref. \cite{hr101}, shows the common domains
of applicability for the LG and the CG as the shaded area, referred to in the
text as the \textquotedblleft oasis\textquotedblright. The line labeled
$\beta_{0}=1$ corresponds to the onset of magnetic field effects at lower
frequencies, as given by Eq. (\ref{o2}). The line labeled $\omega=1/2$ denotes
an upper frequency limit set by the applicability of tunneling methods. A less
restrictive condition for the upper frequency limit is given by Eq.
(\ref{o1}).}%
\label{f1}%
\end{center}
\end{figure}

An essential property of the LG is that it requires only a scalar potential,
as shown in Eq. (\ref{a}). It is this simplicity that makes it so attractive
to use. An immediate inference is that the magnetic field is absent. The
single Maxwell equation required for the LG is Eq. (\ref{r}). Even when
$\mathbf{E}$ has some time dependence, so that $\mathbf{E}=\mathbf{E}\left(
t\right)  $, it remains possible to neglect the magnetic field over a
significant range of laser field frequencies and intensities. This range is
illustrated in Fig. \ref{f1}, taken from Ref. \cite{hr101}. The upper
frequency limit in Fig. \ref{f1} actually refers to tunneling theories rather
than to the less restrictive Eq. (\ref{o1}), but most analytical strong-field
approximations that employ the LG also employ tunneling concepts. This will
not be the subject of further comment since it is the lower frequency limit on
the applicability of the dipole approximation that is often neglected and is
of greater interest in laser applications. The lower frequency limit
(\ref{o2}) on the applicability of the dipole approximation is determined by
the onset of magnetic field effects on a free electron\cite{hr101,hrrev}.%

Following a criterion set some years ago\cite{popov,grbkch}, it has become a
widely accepted practice to judge the validity and reliability of analytical
approximations in plane-wave phenomena by comparing their behavior at zero
frequency with known constant electric field results. \textit{This criterion
is not applicable to plane waves.} Plane waves do not have a zero frequency
limit because that would entail a divergent ponderomotive energy. This is
discussed below.

The shaded area in Fig. \ref{f1} is where dipole-approximation results can be
employed to describe plane wave phenomena (at least for linear polarization).
This can be dubbed the \textquotedblleft oasis\textquotedblright, where
dipole-approximation theories for the description of laser-induced phenomena
can be accurate. A fundamental difficulty is that, starting at any point
within the oasis, there is no path by which zero frequency can be approached
without crossing the lower-frequency limit of the oasis, and entering into a
domain where LG approaches to laser phenomena are not applicable. This
inference is clear, since zero frequency implies the first entry in the chart
(\ref{f}), which is not descriptive of plane wave fields at all. Another way
to understand this is to consider the behavior of the ponderomotive potential
$U_{p}$, which is a fundamental property of plane wave
fields\cite{kibble,bucks,hrjmo,schwinger}. The ponderomotive potential depends
on frequency as%
\begin{equation}
U_{p}\sim\omega^{-2}, \label{t}%
\end{equation}
which means that it is not possible to pass to a zero frequency for a plane wave.

It is often explicitly stated that the zero frequency limit of a plane wave
field exists, and that it is a static electric field. That statement is
motivated by LG concepts. It is not correct. As one of many examples to be
found in the strong-field literature, a recent paper states\cite{arissian}:
\textquotedblleft In adiabatic tunneling the laser field is treated as if it
were a static field ... It is rigorously valid for long wavelengths
...\textquotedblright. This quote starts by citing correctly the essential
property of the LG as treating the laser field as if it were a static electric
field or a QSE field. However, the remark that it is \textquotedblleft
rigorously valid\textquotedblright\ for long wavelengths misidentifies the
limit of a QSE field as $\omega\rightarrow0$ as \ being a legitimate limit of
a plane wave field. The same source continues with the statement:
\textquotedblleft Nonadiabatic tunneling refers to deviations that arise at
higher frequencies ...\textquotedblright, thus retaining the inadequate 1931
view of G\"{o}ppert-Mayer\cite{gm} that the dipole approximation possesses
only a high frequency limit.

\subsection{Ponderomotive potential}

A fundamental property of a charged particle in a laser field is that it
exhibits a ponderomotive potential $U_{p}$ due to the presence of the field.
See Sections 6.2.2 and 7 of \cite{hrjmo}, and Section IV of \cite{schwinger}
for the critical role that $U_{p}$ plays in distinguishing transverse fields
from other types of electromagnetic phenomena. The ponderomotive potential
depends on $A^{\mu}A_{\mu}$ in general, or on $\mathbf{A}^{2}$ in the CG. In
most contemporary laser experiments the dipole approximation applies, which
makes $\mathbf{A}^{2}$ appear to depend only on the time: $\mathbf{A}%
^{2}=\mathbf{A}^{2}\left(  t\right)  $. A basic theorem in mechanics asserts
that any term in the Hamiltonian or Lagrangian function that is a function
only of the time can be removed without any effect on the equations of motion.
However, $U_{p}$ can be very large in practical examples where the dipole
approximation seems to be valid by most criteria, but where the elimination of
$\mathbf{A}^{2}\left(  t\right)  $ would make a huge difference were it not
retained. For example, at a wavelength of $800$ $nm$ and an intensity of
$10^{15}$ $W/cm^{2}$, $U_{p}$ amounts to about $60$ $eV$. The ponderomotive
potential is an essential aspect of plane-wave phenomena, and eliminating it
would make a drastic and unphysical alteration in the description of the
effects of laser fields. The question of the neglect or retention of
$\mathbf{A}^{2}\left(  t\right)  $ has been debated for many years. It led to
the recommendation\cite{hr90} that all strong field problems should be
formulated in relativistic terms on the grounds that strong laser fields make
the field a dominant aspect of a problem. Photon fields propagate with the
velocity of light, so that they impose a relativistic context upon the entire
problem. Starting with a relativistic formalism and then passing to a dipole
approximation when it is warranted should resolve such difficulties as the
retention of $\mathbf{A}^{2}\left(  t\right)  $. That suggestion\cite{hr90}
had little effect on the strong-field community, and the current proposal to
rely on the need to sustain plane-wave properties such as the relativistic
invariants like $F^{\mu\nu}F_{\mu\nu}$ and the onset of explicit magnetic
field effects is a more robust approach.

\section{The \textquotedblleft simpleman\textquotedblright\ theory}

The simpleman theory is so called because it envisions elementary classical
behavior of a charged particle in an electric field as the underlying
explanation for many laser-caused phenomena. A clear statement of this
approach is contained in Ref. \cite{corkum}. At the heart of the idea is the
set of classical solutions for an electron in a sinusoidally varying electric
field. These solutions correspond to the Maxwell equation (\ref{r}). That is,
they are the result of the virtual current $J_{v}$.

\subsection{Linear polarization}

The case of linear polarization presents the opportunity to compare the
predictions of the LG and the CG. An important application of the simpleman
method is for the explanation of high harmonic generation\textit{ }(HHG). For
linear polarization of the laser, the scenario envisioned in the LG is that
atomic ionization occurs by a tunneling process and, after exiting the tunnel,
the photoelectron has its subsequent motion governed by the classical
solutions of (\ref{r}). The motion is oscillatory and primarily along the
direction of the electric field, so that upon reversal of the field direction
the electron is driven back to the parent atom. Recombination can then occur,
with the emission of a photon representing the initial binding energy plus an
additional energy corresponding to a multiple of the frequency $\omega$ of the
driving field.

An explanation in terms of the CG would be similar. The primary difference is
that, in the LG, the electron is \textquotedblleft driven\textquotedblright%
\ by the applied field, whereas, in the CG the laser field is viewed properly
as a transverse field that cannot transfer net energy to a free electron, even
though there is a regular sharing of energy between the field and the charged
particle during any cycle. An equivalent way to state this is to remark that
the LG treats the laser field as if it were a quasistatic electric field
rather than as a proper plane-wave field. The difference is most easily
understood as a contrast between the \textquotedblleft constant electric
field\textquotedblright\ and the \textquotedblleft plane wave
field\textquotedblright\ entries in the chart (\ref{f}). In a pure plane-wave
situation (that is, with no residual Coulomb effects), the trajectory followed
by an electron would be determined entirely by the initial conditions of its
entry into the field. Such trajectories are well known. The most complete
source for this information is probably the article by Sarachik and
Schappert\cite{ss}. The results given there are fully relativistic because a
plane wave field is a relativistic object. However, if the nonrelativistic
limit is taken of the trajectory information given in Ref. \cite{ss}, then the
electron motion is a simple oscillation with\ exactly the same amplitude in
the electric field direction as is found for the LG case. The relativistic
plane wave solution has the well-known \textquotedblleft
figure-8\textquotedblright\ configuration in the frame of reference in which
the electron is at rest on the average over a full cycle, with the long axis
in the direction of the electric field and the short axis in the propagation
direction of the field. In the nonrelativistic limit, the amplitude in the
propagation direction approaches zero, and the LG and CG predictions for the
classical orbit of the free electron become identical.

\subsection{Circular polarization in the LG}

\begin{figure}
[ptb]
\begin{center}
\includegraphics[
height=4.6709in,
width=6.0537in
]%
{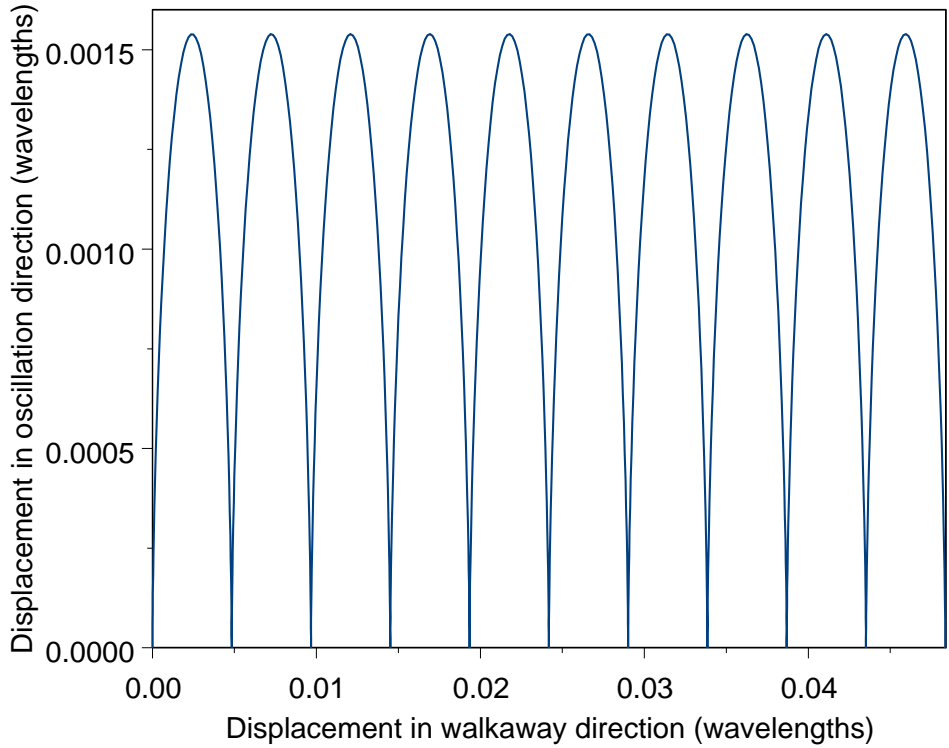}%
\caption{This figure shows the LG particle trajectory as a photoelectron
\textquotedblleft walks away\textquotedblright\ from the atom after ionization
by a circularly polarized field. Ten constant-intensity cycles are shown. The
\textquotedblleft walkaway\textquotedblright\ phenomenon is found only in the
LG, and is a result of the tunneling view of ionization that is appropriate in
the LG, but not in the CG.}%
\label{f2}%
\end{center}
\end{figure}

\begin{figure}
[ptb]
\begin{center}
\includegraphics[
height=4.6709in,
width=6.0537in
]%
{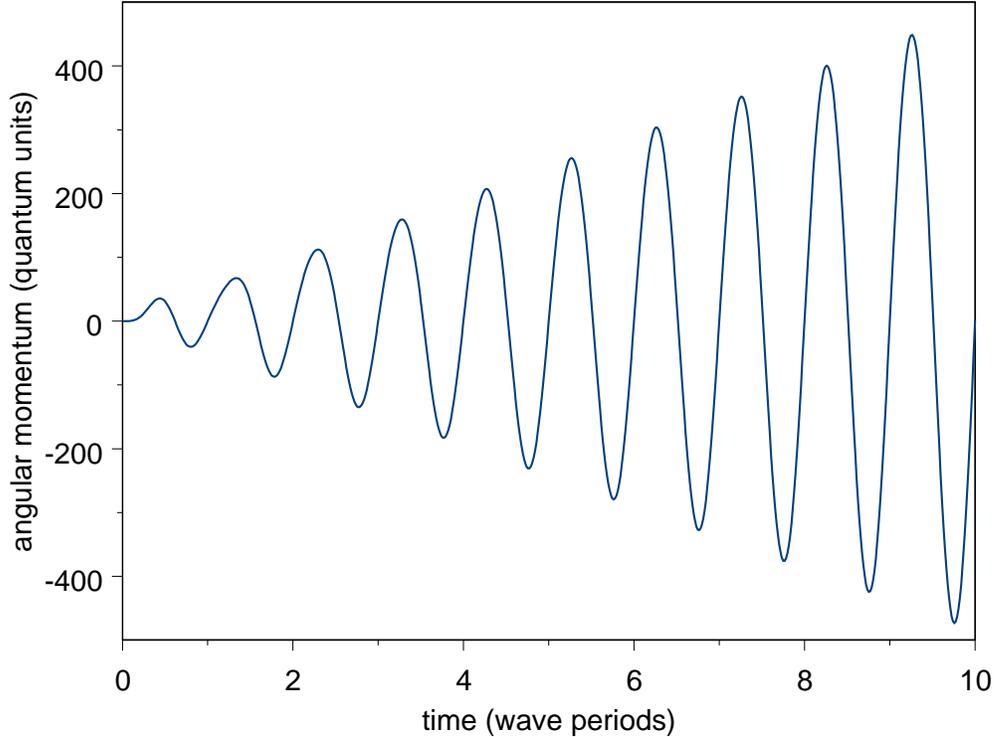}%
\caption{The angular momentum associated with the \textquotedblleft
walkaway\textquotedblright\ phenomenon of Fig. \ref{f2} is shown here. The
angular momentum is calculated with respect to the atom location as the
center. This oscillatory behavior is physically impossible, since\ a
circularly polarized laser can impart only a single sense of rotation.}%
\label{f3}%
\end{center}
\end{figure}

For circular polarization, the LG and the CG give very different results. The
LG case is considered first. Equations (2) and (3) of Ref. \cite{corkum}
provide the necessary information for the classical trajectory of an electron
that exits an atom by tunneling ionization. This sets initial conditions at
zero velocity at the exit of the tunnel, which can be taken to be at the
origin of coordinates. The trajectory is then found to be%
\begin{align}
x  &  =x_{0}\left(  1-\cos\omega t\right)  ,\label{t1}\\
y  &  =x_{0}\left(  \omega t-\sin\omega t\right)  ,\label{t2}\\
x_{0}  &  =\mathcal{E}_{0}/\omega^{2}, \label{t3}%
\end{align}
for a circularly polarized field about the $z$ direction as the axis. The
quantity $\mathcal{E}_{0}$ is the amplitude of the electric field of frequency
$\omega.$ The trajectory is plotted in Fig. \ref{f2} for motion in an electric
field corresponding to a wavelength of $800$ $nm$ at $10^{14}$ $W/cm^{2}$. It
shows an oscillation of the electron of constant amplitude in one direction in
the plane of motion, but with a trajectory in the other direction in the plane
of motion that has been described as \textquotedblleft walking
away\textquotedblright\ from the ion. The most interesting aspect of this
motion comes from an examination of the angular momentum of the electron as
measured from the position of the residual ion. The angular momentum is simply%
\begin{align}
l  &  =xv_{y}-yv_{x},\nonumber\\
&  =\omega x_{0}^{2}\left[  \omega t\sin\omega t-2\cos\omega t\left(
1-\cos\omega t\right)  \right]  , \label{u}%
\end{align}
as follows from Eqs. (\ref{t1}) and (\ref{t2}). This is shown in Fig.
\ref{f3}. After a few cycles, the angular momentum is dominated by the first
term in Eq. (\ref{u}), giving%
\begin{equation}
l\approx\omega^{2}x_{0}^{2}t\sin\omega t, \label{v}%
\end{equation}
a form that makes explicit how the amplitude of the angular momentum
oscillation increases linearly with time. Figures \ref{f2} and \ref{f3} plot
the motion for ten cycles at constant amplitude.%
%

The curious feature in Fig. \ref{f3} is the continuous buildup of angular
momentum, following the behavior shown in Eq. (\ref{v}). In a real ionization
event generated by a circularly polarized laser field (when there is no
difference between the angular momentum states of the initial atom and the
remnant ion), the post-ionization angular momentum should simply be the
angular momentum contained in the number of photons required to ionize. For
the stated parameters, the ponderomotive energy $U_{p}$ of the electron in the
field is about $0.22$ $a.u.$ If the atom being ionized is ground-state
hydrogen, the minimum number of photons required for ionization is%
\begin{equation}
n_{0}=\left\lceil \frac{E_{B}+U_{p}}{\omega}\right\rceil =\left\lceil
\frac{.5+.22}{.057}\right\rceil =13, \label{w}%
\end{equation}
and the most probable kinetic energy of a photoelectron after ionization by a
circularly polarized laser field is a further $U_{p}.$ The most probable
number of photons absorbed for ionization is thus%
\begin{equation}
n=\left\lceil \frac{E_{B}+2U_{p}}{\omega}\right\rceil =\left\lceil
\frac{.5+.44}{.057}\right\rceil =17. \label{x}%
\end{equation}
The bracket symbol that appears in Eqs. (\ref{w}) and (\ref{x}) is the
\textit{ceiling function}, meaning the smallest integer containing the
quantity within the bracket. Each circularly polarized photon possesses one
quantum unit of angular momentum, and all of these angular momenta are
aligned, meaning that the photoelectron should possess about $17$ quantum
units of angular momentum.

Figure \ref{f3} shows a continuous increase of angular momentum, reaching more
than $400\hbar$ after ten cycles. Not only is this a completely unrealistic
magnitude, but the figure shows that the angular momentum oscillates between
positive and negative values. This is not a physically possible outcome, since
a circularly polarized field has a unique sense of rotation and can impart
only one sign of angular momentum to the photoelectron.

\subsection{Circular polarization in the CG}

The relevant Maxwell equations in the CG are Eqs. (\ref{l}) - (\ref{o}). There
are no external sources. In particular, there is no virtual current density
$J_{v}.$ A photoelectron will continue the trajectory which is imparted as the
end result of ionization. This must therefore be a circular orbit centered on
the remnant ion and possessing the energy and angular momentum of a classical
orbit. The means by which this can happen are strongly suggested by a U-matrix
calculation of the ionization of ground-state hydrogen by a single cycle of a
circularly polarized laser field\cite{hrjmo0}. The U matrix tracks the
progress of a particle with time, and approaches the S matrix for large times.
The result of this single-cycle calculation is that the probability density
for the electron moves out to the classical orbit position as the field
amplitude increases. As the amplitude of the field subsequently decreases,
some of the electron probability distribution returns to the distribution of
the neutral atom, but a portion of it is missing. This is the portion that has
been ionized.

A similar scenario, but in the long-pulse regime, is suggested by the
calculation of a correction to the SFA for ionization by circularly polarized
light\cite{hrvk}. The SFA regards the final state of the electron as being
that of a Volkov electron; that is, an electron moving under the influence of
the plane-wave field. The field is regarded as the dominant influence on the
free electron, with a neglect of residual effects of the Coulomb field of the
atom. The correction turns out to be the influence of the Coulomb field on an
electron circulating around the ion in a circular path of classical radius in
the background field.

The recent matching\cite{hr13} of a fully relativistic theory to
experiment\cite{smeenk} confirms the picture of a photoelectron created by
circularly polarized light entering into a circular orbit around the ion.

The underlying reason for both the long-pulse and short-pulse scenarios comes
from the fact that a free electron cannot exchange net energy with a
plane-wave field. During any single cycle, there is an exchange of energy
between the electron and the field but, if the field intensity is maintained
constant, subsequent cycles replicate the cycle history exactly, with no gain
in overall energy. The final state of the photoelectron must thus represent a
photoelectron that satisfies all necessary conservation conditions, including
the conservation of angular momentum that is so glaringly violated in the LG
approach to ionization by a circularly polarized field. The LG acquires from
the virtual current density $J_{v}$ the energy and angular momentum required
to provide the contrasts in Fig. \ref{f3}.%

\begin{figure}
[ptb]
\begin{center}
\includegraphics[
height=4.6709in,
width=6.0537in
]%
{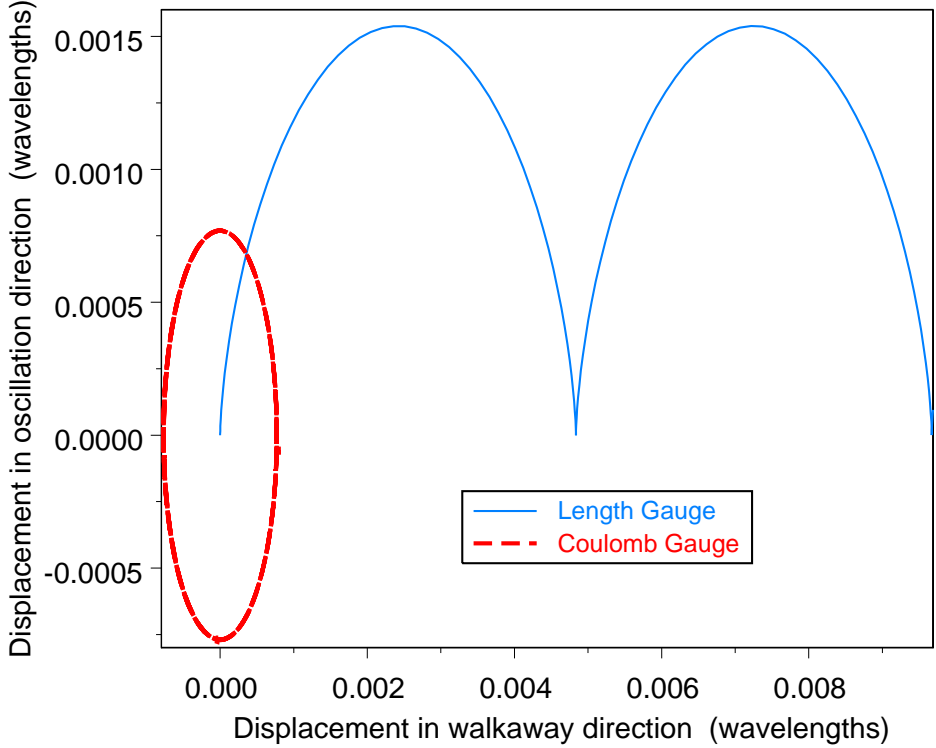}%
\caption{The LG trajectory of the photoelectron shown here corresponds to the
first two of the ten cycles illustrated in Fig. \ref{f2}. The CG trajectory is
a closed orbit around the remnant ion. The scales along the two axes are
different. Were they to the same scale, the CG trajectory would be a circular
orbit around the ion with energy and angular momentum corresponding to
classical parameters for the circular polarization in this example.}%
\label{f4}%
\end{center}
\end{figure}
%

\begin{figure}
[ptb]
\begin{center}
\includegraphics[
height=4.6709in,
width=6.0537in
]%
{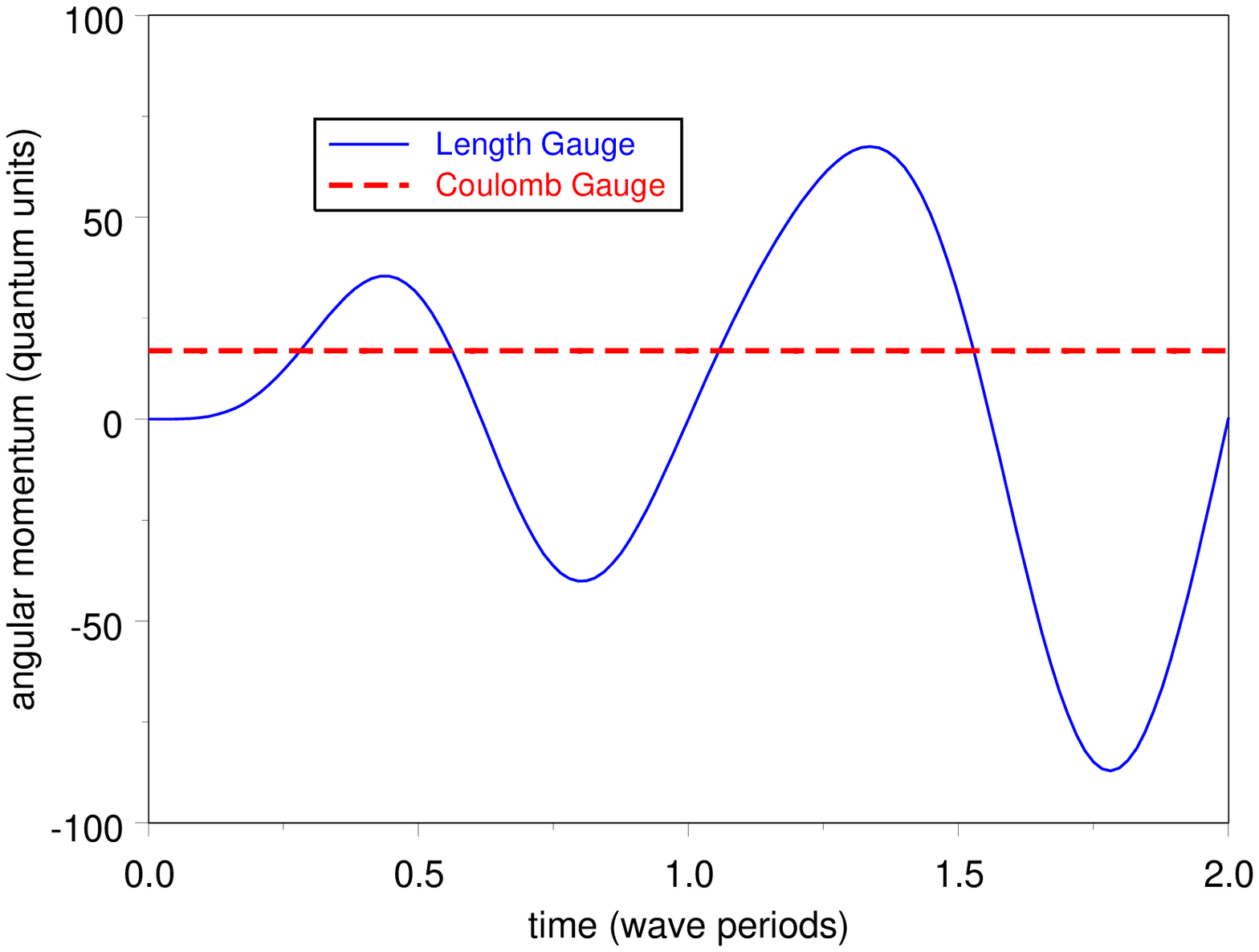}%
\caption{As in Fig. \ref{f4}, only the LG angular momentum history for the
first two cycles of Fig.\ref{f3} are shown here. The CG angular momentum is
constant, so it appears here as a simple straight line at the value given in
Eq. (\ref{x}).}%
\label{f5}%
\end{center}
\end{figure}

Figures \ref{f4} and \ref{f5} relate to the first two of the ten cycles shown
in Figs. \ref{f2} and \ref{f3}. Figure \ref{f4} compares the LG with the CG
prediction for the motion of the photoelectron. Were the two axes scaled the
same, then the CG motion would be circular. Figure \ref{f5} shows the angular
momentum in the CG and LG cases for the same two cycles. The CG - LG contrasts
in Figs. \ref{f4} and \ref{f5} are extreme.

\subsection{Observed qualitative difference between LG and CG for circular
polarization}

The \textquotedblleft walkaway\textquotedblright\ phenomenon discussed above
follows from the general tunneling behavior that the direction of the electric
field is dominant in the ionization process. With circular polarization, the
electric field is always in the radial direction. In the CG, the long-pulse
case simply leads to cylindrical symmetry about the propagation direction, but
the single-cycle U-matrix calculation\cite{hrjmo0} is more explicit. There it
is found that it is the vector potential $\mathbf{A}$ that is the dominant
influence, and the vector potential is $\pi/2$ out of phase with the electric
field. That is, when the electric field is in the radial direction, the vector
potential is in the azimuthal direction. Thus the CG is consistent\cite{hr76}
with the observed \textquotedblleft donut\textquotedblright\ shape of the
photoelectron angular distribution arising from ionization\ by circularly
polarized light. See Fig. 1 of Ref. \cite{bergues}, as discussed in Ref.
\cite{hr76}.

These same properties of photoelectrons arising from circularly polarized
laser light are also confirmed\cite{hr13} in the recent experiments by Smeenk
\textit{et al.}\cite{smeenk}. That is, when ionized by circularly polarized
light, the photoelectron is emitted in a direction perpendicular to the
direction of the electric field, not aligned with the electric field as
implied by the LG.

\section{\textquotedblleft Vector potential\textquotedblright\ in the length
gauge}

The 1980 theory\cite{hr80} of the present author is the only SFA theory that
is formulated in the CG, with subsequent reduction to a dipole-approximation
theory\cite{hrrev,hr90}. Other putative versions of the SFA (including the
Keldysh theory\cite{keldysh}, despite the injunction against naming any
tunneling theory the \textquotedblleft SFA\textquotedblright\ as explained in
Ref. \cite{hr90}) employ Volkov solutions in the LG. This is a curious
situation. There is no possible propagating-wave solution of the Maxwell
equations that can be expressed by a scalar potential alone, yet that is the
defining feature of the LG. The ostensible LG Volkov solution is formulated in
the velocity gauge and then gauge-transformed to the LG. (The velocity gauge
can be derived from the CG by employing dipole-approximation assumptions that
do not alter the Maxwell equations\cite{hrrev,hr90}. Not all velocity-gauge
formulations obey that stricture.) The difficulty is that the Volkov solution
itself (as well as the generating function of the gauge transformation) is
expressed in terms of a vector potential. There is no vector potential in the
LG. For formal purposes, it is envisioned that the dipole-approximation vector
potential is really defined in terms of the electric field through the
relation%
\begin{equation}
\mathbf{A}\left(  t\right)  =-\frac{1}{c}\int_{-\infty}^{t}\mathbf{E}\left(
t^{\prime}\right)  dt^{\prime}. \label{y}%
\end{equation}
This makes the theory nonlocal. The vector potential at any time $t$ depends
on the values of the electric field at all earlier times $t^{\prime}$. This
nonlocality is the price that must be paid to employ the Volkov solution in a
gauge that has no Volkov solution.

The approximation involved in neglecting the final-state Coulomb influence in
one gauge is only indirectly related to making a similar assumption in the
other gauge. The CG and the LG are based on different Maxwell equations, and
the Volkov solution as expressed by a well-defined vector potential in one
gauge is replaced by a nonlocal expression in the other gauge. Therefore, it
is inappropriate to regard the SFA as a theory that lacks gauge invariance; a
statement that occurs frequently in the strong-field literature. The SFA in
the CG and the so-called SFA in the LG are two distinct approximations, and
are not gauge-equivalent versions of the same approximation.

\section{Summary}

Plane waves are transverse fields for which the Lorentz invariant
$\mathbf{E}^{2}-\mathbf{B}^{2}=0$ is a necessity. Any theory for which
$\mathbf{E=E}\left(  t\right)  $ and $\mathbf{B}=0$ are \textit{a priori}
conditions is a theory that can describe longitudinal fields, but not
transverse fields. Within the CG, it is possible to express a
dipole-approximation result, but only when the spatial dependence of the
plane-wave phase is demonstrably unimportant. The LG as applied to plane-wave
problems is not a dipole approximation at all; rather, it amounts to the
assumption that a transverse field can be approximated by a longitudinal
field. The GM equivalence is not complete because the price paid for having a
plane-wave field approximated by a quasistatic electric field is that the
relevant Maxwell equations are quite different in the LG than they are in the
CG. In particular, the LG must possess a virtual charge current density
$J_{v}$ that exists only for the purpose of permitting a nominal gauge
equivalence to exist between a longitudinal field and an approximated
transverse field. This virtual charge current $J_{v}$ does not have
discernible effects under many circumstances, but when an ionization is caused
by a circularly polarized field, the tunneling mechanism (possible in the LG
but not in the CG), requires a set of initial conditions for a photoelectron
that demands major inputs from $J_{v}$. These effects are completely different
from anything in the CG case, and are clearly unphysical for plane waves.

The \textquotedblleft pathology\textquotedblright\ involved in the GM gauge
transformation manifests itself in several ways, some of which have just been
recounted. Another manifestation is in the generating function of the gauge
transformation. This function is proportional to $\mathbf{r\cdot A}$, where
the vector potential $\mathbf{A}$ exists only in the CG. Thus the gauge
transformation, by introducing an alien element into the LG, requires some
nonlocal device such as exhibited in Eq. (\ref{y}). This is just another of
the several symptoms of the pathology.

These fundamental differences between the LG and the CG have implications for
the SFA. They mean that the approximation employed in the SFA has different
consequences in the two gauges. The statement that can be found in the
literature that the SFA is somehow defective because it is gauge dependent, is
a \textit{non sequitur}.

A simple way to summarize what has been said here is the following: The LG
provides a description of quasistatic electric field phenomena, because that
is all that can be described by a scalar potential alone. That is, the LG
represents the first line in the chart (\ref{f}). Plane wave phenomena require
a vector potential, since that is necessary for the description of a
transverse field or, equivalently, the description of a propagating
electromagnetic field. The second entry in the chart (\ref{f}) relates to this
phenomenon. These are the basic differences that underlie the various problems
discussed above. What has confused the situation is the assumption that Eq.
(\ref{1a}) is an unqualified condition for both the LG and the
dipole-approximated CG. This underlies the original premise of the
G\"{o}ppert-Mayer gauge transformation. G\"{o}ppert-Mayer cannot be faulted
for not seeing this problem in 1931. Perturbative AMO physics has thrived for
many decades on the LG. However, Eq. (\ref{1a}) is not universally compatible
with plane waves, and this is an important distinction in strong-field,
nonperturbative physics.

\end{document}